\documentclass[aps,amsmath,amssymb,prb,twocolumn,superscriptaddress]{revtex4}

\usepackage{amsmath}
\usepackage{graphicx}
\usepackage{amssymb}
\usepackage{amsthm, amscd}
\usepackage{amsfonts}
\usepackage{times}
\usepackage{pstricks}
\usepackage{wasysym}

\begin{document}

\title{Exact Solution for Bulk-Edge Coupling in the Non-Abelian $\nu=5/2$ Quantum Hall Interferometer}

\author{ Bernd Rosenow}
\affiliation{Max-Planck-Institute for Solid State Research, Heisenbergstr. 1, D-70569 Stuttgart, Germany }
\author{ Bertrand I. Halperin}
\affiliation{ Physics Department, Harvard University, Cambridge
02138, Massachusetts, USA}
\author{Steven H. Simon}
\affiliation{Rudolf Peierls Centre for Theoretical Physics, 1 Keble Road, Oxford, OX1 3NP, UK}
\author{Ady Stern}
\affiliation{ Department of Condensed Matter Physics, Weizmann
Institute of Science, Rehovot 76100, Israel}

\date{June 1, 2009}

\begin{abstract}
It has been predicted that the phase sensitive part of the current through a
non-abelian $\nu = 5/2$ quantum Hall Fabry-Perot interferometer  will
depend on the number of localized charged $e/4$ quasiparticles (QPs) inside the
interferometer cell.
In the limit where all QPs are far from the edge, the
leading contribution to the interference current is predicted to be absent if 
the number of enclosed QPs is odd and present otherwise, as a consequence of
the non-abelian QP statistics.
The situation is more complicated, however,  if a localized QP is  close enough
to the boundary so that it can exchange a Majorana  fermion with the edge
via a tunneling process.
Here, we derive an exact solution for the dependence of the interference current
on the  coupling strength for this tunneling process, and  confirm a previous
prediction that for sufficiently strong coupling, the localized QP is
effectively incorporated in the edge and no longer affects the interference
pattern.  We confirm that  the dimensionless  coupling strength  can be tuned
by the source-drain voltage, and we find that  not only does the magnitude of
the even-odd effect change with the strength of bulk-edge coupling, but in
addition, there is a universal shift in the interference phase as a function of
coupling strength. Some implications for experiments are discussed at the end.

\end{abstract}


\maketitle

\section{Introduction}

Quantum mechanical systems with  topological excitations are, in principle,   ideally qualified  for quantum information 
processing  as   the state of the topological sector  is not accessible to local operators, and hence local perturbations cannot lead to decoherence. The best studied example of topologically ordered states are fractional quantum Hall (QH) systems, where  braiding of quasiparticles changes the ground state wave function. 
While in  conventional QH 
 states with odd denominator filling fraction only the phase of the wave function is changed by quasiparticle (QP)  braids, 
the more recently discovered quantum Hall state at filling fraction 5/2 is expected to support non-abelian QPs, whose 
braiding corresponds to transformations in a  degenerate ground state manifold.  \cite{review,sternreview}  The ground state degeneracy can be used to store information in the form  of quantum bits, 
with one qubit for a pair of QPs, and 
quantum gate operations may be performed by the braiding  of  QPs.
In principle, the read-out of quantum bits is possible by means of interference experiments. 
\cite{dassarma05,stern06,bonderson06}

One possible device for the readout of a topological quantum bit is the Fabry-Perot interferometer. It consists of two 
narrow constrictions in a Hall bar which act as quantum point contacts and allow backscattering between counter-propagating edges modes. Interference between partial waves backscattered at the first and at the second quantum point contact is sensitive to the phase acquired during a trip around the interferometer cell. Proposals for using a Fabry-Perot interferometer for the readout of a topological quantum bit rely on the fact 
that there is a relative change of the  interference phase by $\pi$ depending on the state of the qubit enclosed in the cell. \cite{dassarma05,stern06,bonderson06}
 
The 5/2 quantum Hall state can be described as a p-wave superconductor of composite fermions. \cite{readgreen} In this picture, QP excitations are vortices accompanied by an electric charge $\pm e/4$. These vortices have a zero energy bound state at their core, which is described by a Majorana degree of freedom. The Majorana operators 
associated with two such vortices can be combined into a complex fermion
and constitute a two level system suitable for 
quantum information processing. The state of this two level system can be changed by moving a third QP around one 
partner of the pair. \cite{NaWi96}  Depending on the occupancy  of this two level system, the interference phase obtained by a partial wave that encircles the two level system  is predicted to change by $\pi$. More generally, the influence of bulk particles localized inside the interference cell on the Fabry-Perot interference phase can be used to provide evidence for the non-abelian character of 5/2-QP excitations.

As an example, we may consider the phase change  when  voltage is applied to an ideal side gate, which is able to vary the area $A$ enclosed by the interferometer path, without changing the electron density inside.   If the interference signal is caused by backscattering of $e/4$ quasiparticles at the constrictions of the interferometer, and if there are no localized QPs inside the loop, then the phase of the interferometer signal will change by $2 \pi$ when the area is varied by an amount 
$\Delta A = 4 \Phi_0 / B$, where $B$ is the magnetic field strength and $\Phi_0$ is the flux quantum for an electron. 
For an odd number of bulk QPs inside the interferometer cell, however, the leading sinusoidal dependence of the interference current on this  phase is expected to vanish, while it is restored for an even number of bulk QPs.\cite{dassarma05,stern06,bonderson06} This dependence of the interference signal on the parity of bulk QPs in the interferometer cell constitutes the so-called even-odd effect. 

In order for the even-odd effect to be observable, it is necessary that the quantum state of the localized QPs remain independent of time during the course of the current measurement.  This can be a problem, in real systems, because of tunnel coupling between the bulk QPs and the edge.
In principle, one can imagine two 
types of tunnel couplings: tunneling of charged $e/4$ QPs into and out of the interferometer cell
\cite{GSS06}, and coupling between bulk Majorana degrees of freedom and the Majorana mode along the edge. The former process should be suppressed by 
the requirement of charge neutrality due to the Coulomb interaction, which is expected to be strong in small interference devices.  The latter process is likely to be experimentally relevant, and it is this process which is the focus of the current paper.

 As tunnel matrix elements 
typically depend on distance in an exponential way, QPs localized near the sample edge  can  have a sizable tunnel coupling. In device geometries defined by etch trenches or top gates, the electron density typically
changes from its maximum value to zero over a distance of many magnetic lengths, such that one can expect  the filling factor to deviate from the ideal value of 5/2  in some region near the interfering edge states. However, any deviation from  
the ideal filling fraction  5/2 implies a finite density of QPs near the edge. Due to the spatial proximity to the interfering edge states, 
these localized QPs can have a significant tunnel coupling to the edge, and 
a  realistic description of a non-abelian Fabry-Perot 
interferometer needs to take bulk-edge coupling into account.  

The case of one weakly  coupled  bulk QP was analyzed in perturbation theory by Overbosch and Wen \cite{overbosch07}, and the case of two bulk QPs coupled to opposite edges was studied by the present authors. \cite{RoHaSiSt08} While for weak coupling,  an analytic perturbative solution was possible, the strong coupling regime 
was analyzed numerically for a lattice model.   
It was found that at $T=0$, the dimensionless
parameter describing the strength of the  bulk-edge coupling may be written as
$ \hbar \lambda^2/ v_n |e^*V |$, where $\lambda$ is the tunnel-coupling
strength, $v_n$ is the velocity of the Majorana edge mode, $V$  is the
source-drain voltage, and $e^*=e/4$ is the QP charge.
For small values of this parameter, bulk QPs effectively decouple from the edge, while for a large coupling strength they are absorbed by the edge. In this manuscript, we present an exact solution for the influence of 
bulk-edge coupling on the magnitude and phase  of interference in a non-abelian Fabry-Perot interferometer.

For example, we analyze the case of a single bulk QP whose Majorana mode is coupled to one edge of the interferometer.
We  find that the  interference current can be exactly evaluated in the regime where the length $b$ 
of the interferometer is short compared to  $\hbar v_n / e^* V$.  The interference current is reduced
 relative to the case of no bulk impurity by the modulus of 
%
\begin{equation}
J_{\rm imp}\left({\hbar \lambda^2  \over v_n |e^\star V|} \right) \ = \ \left( 1 + {i \over 2} { v_n |e^\star V|
\over \hbar \lambda^2} \right)^{- {1 \over 2}} \ \ , 
\label{voltagereduction.eq}
\end{equation}
%
and  the interference phase is shifted by the phase of this expression. 
Thus, in the presence of 
bulk-edge coupling the even-odd effect is modified in an important way: instead of the absence of the
leading harmonic of the interference current, this harmonic grows with decreasing voltage at a rate which is enhanced relative to the behavior in the absence of a bulk Majorana mode (the intensity grows  $\propto
1/V$ instead of $\propto 1/\sqrt{V}$ in the high voltage regime). In addition, there is a universal phase shift of $e^{i \pi/4}$ in the low 
voltage regime relative to the high voltage regime. 
The result Eq.~(\ref{voltagereduction.eq}) is 
in agreement with previous perturbative and numerical results. The result Eq.~(\ref{voltagereduction.eq}) 
predicts interference contrast and phase for arbitrary coupling strength and should be relevant for the interpretation of $\nu=5/2$ interference experiments.
 
The manuscript is organized as follows: in section II we introduce our model in a continuum formulation; in section III we discuss the implications of our exact solution for interference experiments; in section IV we describe the lattice formulation of the model; in section V we derive the exact solution of the lattice model; and  in section VI we present an interpretation of this 
solution in terms of a resummed perturbation theory. In section VII we conclude with a discussion of our main results and some additional comments on their application to experiments.  While sections I, II, III, and VII are intended for the more general reader, sections IV to VI are more mathematical and describe the derivation of our results in some detail.

\section{Continuum Model}

The model we  consider is a Hall bar  parallel to the $x$-axis  with two quantum point contacts \cite{stern06,bonderson06,overbosch07,RoHaSiSt08,fradkin98,chamon97,BisharaNayak08}, see Fig.~1.
In the absence of any coupling to bulk quasiparticles the upper ($u$)
and lower ($d$) edges of the $\nu=5/2$ state are described by two
charged boson fields $\phi_u(x),\phi_d(x)$ and two neutral Majorana
fermion fields $\psi_{u}(x)$, $\psi_{d}(x)$. The Lagrangian density for the boson field on each edge is that
of a chiral Luttinger liquid with velocity $ v_c$
%
\begin{equation}
{\cal L}_c^r \ = \ {1 \over 4 \pi  }  \partial_x \varphi_{r} ( v_c \partial_x \pm i 
\partial_\tau) \varphi_{r} \ \ .
\end{equation}
Here, $r=u, d$ denotes the upper and lower edge, and the plus sign goes with $r=u$, the minus sign with $r=d$. 
The charge density is given by $\rho_{r} = {1 \over 2 \pi} \partial_x \varphi_{r}$. For simplicity we set $\hbar =1$ 
when no confusion results. 
The Majorana fields encode the non-abelian properties of the 5/2-state, their  Lagrangian densities are 
%
\begin{equation}
{\cal L}_n^r ={1 \over 4 } \psi_{r}(x)( \partial_\tau  \pm v_n \partial_x) \psi_{r} (x) \ \ .
\label{majorana.eq}
\end{equation}
%
Within the p-wave superconductivity picture of the 5/2-state, QPs with a charge $\pm e/4$ are vortices of the superconductor, which have 
Majorana bound states at their core. In our model, there are two localized bulk QPs which carry a zero mode Majorana each, 
described by a localized 
Majorana operator. We denote the two bulk
Majorana operators by $\Gamma_u,\Gamma_d$, with the subscript
indicating the edge to which the quasiparticle couples. 
In the absence of coupling, the Lagrangians of these bulk Majorana modes are
%
\begin{equation}
{\cal L}_b^r  =  { 1 \over 4} \Gamma_{r} \partial_\tau \Gamma_{r} \ \ .
\end{equation}
%
We assume the two bulk QPs to be spatially well separated from each other such that there is no coupling between the Majorana modes $\Gamma_u$ and $\Gamma_d$ associated with them. 
The
two-dimensional Hilbert space created by these Majorana modes is
spanned by the two eigenvectors of the operator $i\Gamma_u\Gamma_d$.
The coupling of $\Gamma_u$ to
the upper edge and $\Gamma_d$ to the lower edge, both at $x=x_0$, 
is described by  the
Lagrangian density 
%
\begin{equation}
{\cal L}_{b-e}=
i\left[\lambda_u\psi_u(x)\Gamma_u+\lambda_d\psi_d(x)\Gamma_d\right
]\delta(x - x_0) \ \ .
 \label{bulk-edge-tunneling}
\end{equation}
%
Bulk-edge coupling gives rise to tunneling times $t_{\lambda r} = (\pi \lambda^2_{r}/2 v_n)^{-1}$. In order 
to judge the effect of bulk-edge coupling on the interference signal, the tunneling time has to be compared to 
the geometric time $t_b = b/v_n$ needed to move between the two constrictions separated by a distance $b$, and to the voltage time 
$t_V = \hbar / e^* V$, which can be interpreted as the extension in time of a QP wave packet. In the limit
$t_V \gg t_b$, in which the interference signal is most clearly seen, the effective strength of bulk-edge coupling is 
given by the ratio $t_V/ t_\lambda$. As we shall see, if this ratio is much smaller than one, the bulk state is effectively decoupled from the edge, and if the ratio is much larger than one, the bulk state is effectively absorbed by the edge and does not influence the interference signal any more. \cite{RoHaSiSt08}

A charge $e/4$ QP consists of a charge part and the neutral Majorana mode associated with it.  
 The operator that tunnels a quasiparticle across a constriction is the product 
of a charge part  and a neutral 
part which encodes the non-abelian statistics of quasiparticles.
The tunneling part of the Hamiltonian  is 
%
\begin{equation}
H_{tun}\equiv {\hat T}
+{\hat T}^\dagger  \ \ , 
\label{tunnelinghamiltonian.eq}
\end{equation}
%
where
%
\begin{equation}
{\hat T}=e^{i e^* V t} \left[ \eta_L  {\cal C}_L{\cal N}_L +\eta_R {\cal
C}_R{\cal N}_R i \Gamma_u \Gamma_d \right]
\label{tunneling-schematic}
\end{equation}
%
transfers a quasiparticle from the lower to the upper edge through the
left $(L)$ and right $(R)$ constrictions respectively, and its
hermitian conjugate ${\hat T}^\dagger$ similarly transfers a quasiparticle
from the upper to the lower edge. Alternatively, one can say that $\hat{T}$ transfers a 
quasihole from the upper to the lower edge, and that ${\hat T}^\dagger$ transfers a 
quasihole from lower to upper edge. The operators ${\cal C}_{L,(R)}$ and 
${\cal N}_{L,(R)}$ will be defined below. 
The Aharonov-Bohm phase is absorbed into the relative phase
between the tunneling coefficients $\eta_{L}$ and $\eta_R$. 
Here, $V$ is the voltage difference
between the two edges, and $e^* = e/4$ is the quasiparticle
charge. Correspondingly, the current operator is given by 
%
\begin{equation}
\hat{J} =
\frac{e^*}{i}(\hat{T} -\hat{ T}^\dagger) \ \ .
\end{equation}
%
 The operators 
%
\begin{equation}  
 {\cal C}_{L(R)}\equiv
e^{i\left (\phi_u(\mp b/2)-\phi_d(\mp b/2)\right )/2}
\label{chargetunnel.eq}
\end{equation}
%
are the
charge part of the tunneling operator, operating on the charge
mode. The factor $1/2$ in the exponential reflects the fact that the QP charge
$e/4$ is one half of the "natural" charge of a state with filling fraction $\nu= 1/2$.

%
\begin{figure}[h]
\includegraphics[width=0.9\linewidth]{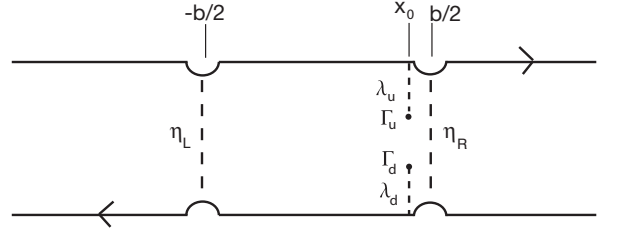}
\caption{Interferometer  with quasi-particle tunneling at positions 
$-b/2$ and $b/2$. One localized Majorana mode couples to  the upper edge at spatial position $x=x_0$, another one to the lower edge at the same  position. For the derivation of the exact solution, the bulk Majoranas are 
positioned at the boundary of the interferometer cell with $x_0=b/2$. }
\label{correlation.fig}
\end{figure}
%

The neutral parts of
the tunneling operators can be expressed as spin fields of an Ising CFT \cite{YellowBook}
%
\begin{equation}
 {\cal N}_{L(R)} \equiv
\sigma_u(\mp b/2)\sigma_d(\mp b/2)  \ \ .  
\label{neutraltunnel.eq}
\end{equation}
%
The $\sigma$
operators can be  defined through their operation on the Majorana
fermion fields as \cite{SBPRB}
%
\begin{equation}
\psi_{r}(y)\sigma_r(x_0)=  \mp {\rm sgn}(x_0-y) \sigma_r(x_0) \psi_{r}(y)
\label{sigmadef}
\end{equation}
%
with $r=u,d$, and the minus sign going with the upper edge. We will discuss an alternative  expression for the neutral tunneling operator in the next paragraph.   The factor of $i \Gamma_u \Gamma_d$ in the second term of
Eq.~(\ref{tunneling-schematic}) is included to account for the
wrapping of a tunneling quasiparticle at position $x=b/2$ around the two
localized quasiparticles.
This factor is responsible for  the $\pi$
phase shift between the interference patterns corresponding to the two
eigenvectors of $i \Gamma_u \Gamma_d$. 

The neutral part of the tunneling operator can be expressed  in a more intuitive way by 
using the parity operator for the part of the system to the left of the tunneling site. We arrive at this formulation by 
representing the 5/2 state as a p-wave superconductor of composite fermions \cite{readgreen}. In this 
picture, the quasiparticle with charge $e/4$ is a vortex of the superconductor. The superconducting phase 
changes by $2 \pi$ when encircling the vortex so that the condensate wave function is single valued.
As  a Cooper pair has two fermions, the phase of the fermionic wave function changes only by $\pi$ when encircling the 
vortex, and there has to be a branch cut in the phase field seen by unpaired fermions
to make their wave function single valued. For this reason, every vortex drags behind it a branch cut in the phase field of unpaired fermions. 
 The phase jump of $\pi$ across the branch cut shows up as the minus sign in 
the commutation relation Eq.~(\ref{sigmadef}): while a Majorana operator at spatial  coordinate $y > x_0$ is  not affected by the tunneling of a charge $e/4$ QP as described by a $\sigma$ operator, for $y < x_0$ the Majorana operator acquires an extra minus sign. 

Alternatively, the minus sign a Majorana operator acquires when crossing the branch cut left behind by an $e/4$ QP can  be described by  an operator which shifts the phase of fermions by $\pi$. 
This operator can be found by using an analogy with spatial translations. A spatial translation by a distance $a$ is 
described by the exponential $\exp(i \hat{p} a)$, where $\hat{p}$ is the momentum conjugate to the spatial coordinate 
which is shifted by $a$. In order to describe a "phase translation", we need the exponential of the operator conjugate to  the phase operator. In a superconductor, phase and  number of cooper pairs are conjugate so that the Cooper pair number operator 
generates phase shifts. At zero temperature, the Cooper pair density is  just half the electron density $\hat{\rho}(x)$, and the operator 
%
\begin{eqnarray}
\hat{P}(- \infty, x_0) &  = & e^{i \pi \hat{N}(- \infty, x_0)}
\label{paritydef_eq} \\ 
& = & e^{i \pi \int_{x < x_0} d^2 r \hat{\rho}(r)} \nonumber 
\end{eqnarray}
%
generates a relative phase shift of $\pi$ between operators $\psi(y)$ with $y < x_0$ and $y > x_0$, respectively. As $ \hat{N}(- \infty, x_0) $ is 
the fermion number operator for the region $x < x_0$, the operator 
$e^{i \pi \hat{N}(- \infty, x_0)}$ is just the parity of the number of fermions to the left of $x_0$, and we identify the 
operator $\hat{P}(- \infty, x_0)$ defined in Eq.~(\ref{paritydef_eq}) as the parity operator. 
When 
evaluating $\hat{P}(- \infty, x_0) \psi_r(y)$ for $y < x_0$, the change in particle number due to the action of $\psi_r(y)$ is included in the evaluation of $\hat{P}(- \infty, x_0)$ and changes its value by minus one, while for the opposite operator order $ \psi_r(y)\hat{P}(- \infty, x_0)$ this is not the case. In this way, Eq.~(\ref{sigmadef}) is reproduced for $y < x_0$ when making the 
identification 
%
\begin{equation}
\sigma_u(x_0) \sigma_d(x_0) = \hat{P}(- \infty, x_0) \ \ .
\end{equation}
%
Clearly, for $y > x_0$ the order of $\hat{P}(- \infty, x_0)$ and $\psi_r(y)$ does not  influence the value of $\hat{P}(- \infty, x_0)$, and Eq.~(\ref{sigmadef}) is reproduced again. 

In the following, it will be  useful to decompose the parity operator into a bulk part $\hat{P}_{\rm bulk}$ measuring the parity of bulk Majoranas and an 
edge part $\hat{P}_{\rm edge}$. 
In order to keep our model simple, we assume that any localized QPs in the region 
$x < - b/2$ are far from the edge, so that the occupation number of their associated 
Majorana states do not change during the course of the experiment. 
Since the parity operator factorizes 
according to 
$\hat{P}_{\rm bulk}(- \infty,b/2) =   \hat{P}_{\rm bulk}(- \infty,-b/2) \hat{P}_{\rm bulk} (- b/2,b/2)$, under the above assumption it 
is sufficient to include only $ \hat{P}_{\rm bulk}(-b/2,b/2)$ in the tunneling operator for the right constriction. 
In our model, there are only 
two bulk Majoranas $\Gamma_u$, $\Gamma_d$ inside the interferometer cell with $-b/2 < x < b/2$. Their parity 
is determined  by the operator $i \Gamma_u \Gamma_d$, which indeed appears as a factor in the definition 
Eq.~(\ref{tunneling-schematic}) of the tunneling operator. 

As the bulk part of the parity operator is fully described by the factor $i \Gamma_u \Gamma_d$ in the tunneling Hamiltonian, the neutral 
operators ${\cal N}_{L,R}$ can be expressed in terms  of the edge parity operator. 
In order to find an explicit expression for it, we express the particle density on upper and lower 
edge together as $\hat{\rho}_{\rm edge}(x) = i \psi_u(x) \psi_d(x)$ and find 
%
\begin{equation}
P_{\rm edge}(-\infty,x_0) = e^{i \pi \int_{-\infty}^{x_0} dx\, i \psi_u \psi_d } \ \ . 
\label{edgeparity.eq}
\end{equation}
%
As the edge parity operator factorizes in the same way as the bulk parity operator, the equal time neutral correlation function $ \langle {\cal N}(-b/2) {\cal N}(b/2)\rangle = \langle  P_{\rm edge}(-b/2,b/2) \rangle$ 
is given by the expectation value of the edge parity operator for the interferometer cell
%
\begin{equation}
 P_{\rm edge}(-b/2,b/2) \ = \   e^{i \pi \int_{-b/2}^{b/2} dx\, i \psi_u \psi_d } \ \ . \label{edgeparity_static.eq}
\end{equation}
%
This expression will be useful for the treatment of the lattice model introduced in section IV. 
In the framework of this  lattice model, the edge parity operator reduces to a product over local parity operators.

\section{Signatures of bulk-edge coupling in the interference current}

To lowest order in the tunnel couplings $\eta_L$, $\eta_R$, the expectation value of the interference contribution to the backscattered current can be obtained using linear response theory. In this approach, the perturbation is the tunneling 
Hamiltonian Eq.~(\ref{tunnelinghamiltonian.eq}) and (\ref{tunneling-schematic}). Starting from the relation 
%
\begin{equation}
\langle I \rangle  \ = \ {1 \over i \hbar} \int_{- \infty}^0 dt \langle  \left[ J(0), H_{\rm tun}(t)\right] \rangle \ \ , 
\end{equation}
we find after some algebra that the interference contribution to the backscattered  current is given by 
%
\begin{eqnarray}
I_{\rm int} & = & {4 e^\star \over \hbar^2} {\rm Re}\  i \eta_L^\star \eta_R \int_{-\infty}^\infty dt e^{-i e^\star V t/\hbar} 
 {\rm Im} \Big[ \langle {\rm T}_\tau {\cal C}_L^\dagger(\tau) {\cal C}_R(0)  \rangle 
\nonumber \\
& & \times \langle  {\rm T}_\tau {\cal N}_L(\tau) {\cal N}_R(0)   i \Gamma_u(0) \Gamma_d(0) \rangle \Big]\Big|_{\tau \to i t + \delta} 
\ \ ,
\label{interferencecurrent.eq}
\end{eqnarray}
%
where $T_\tau$ is the time ordering operator, and 
$\delta$ is the short time cutoff of the theory. 
Using the definition Eq.~(\ref{chargetunnel.eq}), the expectation value of the charged correlator can be directly evaluated as 
%
\begin{eqnarray}
\langle {\rm T}_\tau {\cal C}_L^\dagger(\tau) {\cal C}_R(0)  \rangle  & = &
{\delta^{1/4} \over \big[ \tau^2 + b^2/v_c^2]^{1/8} } \ \ .
\end{eqnarray}
%
In the absence of bulk-edge coupling, the neutral correlator can be obtained from the representation Eq.~(\ref{neutraltunnel.eq}) by using the  the conformal dimension $h_\sigma = {1 \over 16} $ of the 
$\sigma$-field  in the expression for CFT correlation functions \cite{YellowBook}. One finds
%
\begin{eqnarray}
\langle {\rm T}_\tau {\cal N}_L^\dagger(\tau) {\cal N}_R(0)  \rangle \Big|_{\tau \to i t + \delta}&  = &
{\delta^{1/4} \over \big[ \tau^2 + b^2/v_n^2\big]^{1/8}} \ \ . 
\end{eqnarray}
%
Alternatively, it can be calculated by using  a 
bosonized version of Eq.~(\ref{edgeparity.eq}).
Using the lattice model described in the next section, we have been able to obtain an exact solution  for the neutral correlation function in the presence of two impurities at 
$x_0 = {b \over 2}$, one of them coupled to the upper edge and the second coupled to the lower edge. From Eq.~(\ref{reduction_time.eq}) we find
%
\begin{eqnarray}
\langle {\rm T}_\tau {\cal N}_L^\dagger(\tau) {\cal N}_R(0) i \Gamma_u(0) \Gamma_d(0)  \rangle &= & 
{2 \delta^{1\over 4} \over \pi v_n}  
   \left( {b^2\over v_n^2} + \tau^2\right)^{3\over 8} 
\nonumber\\
& & \hspace{-3.5cm} \times {\lambda_u }
\left[ e^{\lambda_u^2(b - i v_n \tau)/v_n^2} K_0 (\lambda_u^2(b - i v_n \tau)/v_n^2)\right]  \nonumber \\
& & \hspace{-3.5cm} \times   {\lambda_d } \left[ e^{\lambda_d^2 (b + i v_n\tau)/v_n^2} K_0\left(\lambda_d^2 (b+ i v_n \tau)/v_n^2 \right) \right] \nonumber \ , \\ \label{twoimpneutral.eq}
\end{eqnarray}
%
where $K_0(x)$ is the modified Bessel function of order zero. 
Using this expression in Eq.~(\ref{interferencecurrent.eq}), the interference current can be evaluated for arbitrary system size, 
bulk edge coupling, and all ratios of $v_n/v_c$. For illustrative purposes, we evaluate now the  interference current in  the regime of small interferometer length $b$, in which the interference contrast is highest and where the size $b$ of the interferometer cell can be set to zero. 
In addition,  we will concentrate on a situation with only one impurity present in the bulk, say $\Gamma_u$.
This situation can be described by sending ${b \lambda_d^2 \over v_n^2} \to \infty$ in Eq.~(\ref{twoimpneutral.eq}) such that the 
impurity degree of freedom $\Gamma_d$ is effectively absorbed by the edge. 
This situation has the benefit of being more easily  interpreted than the two impurity case. We will see that 
the visibility of the interference signal grows from zero to one as the coupling strength is increased.  
We define 
$\omega = - e^\star V / \hbar$ and find 
%
\begin{eqnarray}
I_{\rm int} & = & {4 e^\star \over \hbar^2} {\rm Re}\  i \eta_L^\star \eta_R
\sqrt{\delta} \ {\lambda_u \over \sqrt{v_n}} 
\label{impuritycurrent.eq} \\
& &\hspace{-1cm}  \times  \sqrt{2\over \pi} \int_{-\infty}^\infty dt e^{i \omega t}  
{\rm Im} \left\{
{\sqrt{ t - i \delta} \ e^{\lambda_u^2 t/v_n}  K_0\left[{\lambda_u^2 \over v_n} (t  - i \delta) \right] \over \left[ - t^2 + i \delta\,  t  \right]^{1 \over 4}     }\right\}\nonumber  \\
& = & {4 e^\star \over \hbar^2} \sqrt{\delta \pi \over \omega} \ 
  {\rm Re} \left[  \eta_L^\star \eta_R  J_{\rm imp}\left( {\lambda_u^2 \over \omega v_n}\right) \right]\nonumber
\end{eqnarray}
%
with 
%
\begin{eqnarray}
J_{\rm imp}(x) & = &  {1  \over \sqrt{ 1 + {i \over 2 x}}} \ \ .
\end{eqnarray}
%
To evaluate the time integral in Eq.~(\ref{impuritycurrent.eq}), we split the domain of integration into negative and positive times. In each subdomain, the square roots of time 
in numerator and denominator cancel up to a phase factor so that the integral is a Fourier
integral of Bessel functions, which can be found in the literature \cite{GraRy}. For positive arguments, the Bessel function $K_0(x)$ is real,
 while for negative arguments $x<0$,  $K_0(x)$ has a cut along the real axis and can   be decomposed into real and imaginary part according to 
$K_0(x - i \delta) = K_0(-x) + i \pi I_0(-x)$.

With respect to the case without bulk impurity, the interference signal is modified by the additional factor 
$J_{\rm imp}(\lambda_u^2/\omega v_n)$ in Eq.~(\ref{impuritycurrent.eq}). The modulus of $J_{\rm imp}$ reduces the 
amplitude of interference oscillations for small values of $\lambda_u^2/\omega v_n$, while the argument of $J_{\rm imp}$ gives rise to a phase shift. Both modulus and 
phase of $J_{\rm imp}(x)$ are displayed in Fig.~2. 
%
\begin{figure}[h]
\includegraphics[width=0.9\linewidth]{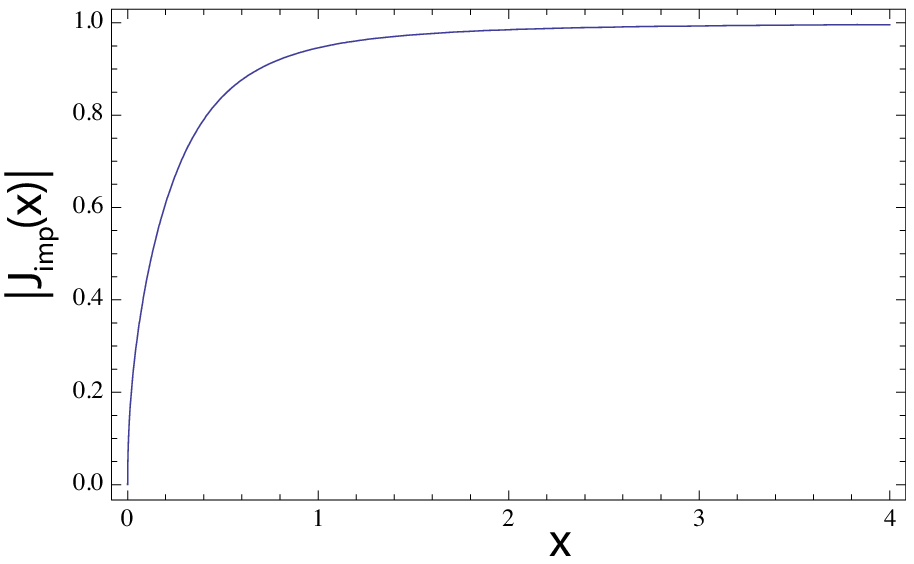}
\includegraphics[width=0.9\linewidth]{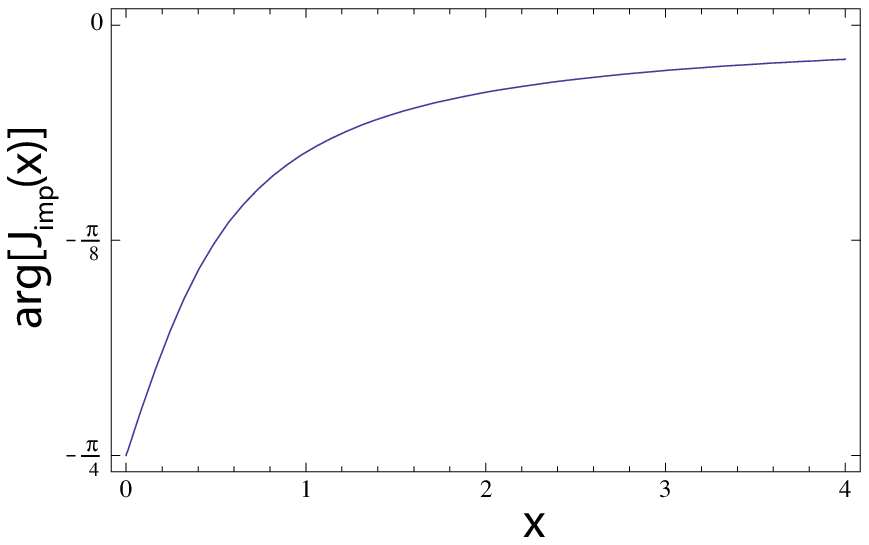}
\caption{Modulus and argument of the factor $J_{\rm imp}({\lambda^2 \hbar \over - v_n e^\star V})$ describing the 
modification of the interference current due to bulk-edge coupling. }
\label{Jimp.fig}
\end{figure}
%
The expansion of $J_{\rm imp}(x)$  for small arguments (weak tunneling or large voltage) is 
%
\begin{equation}
J_{\rm imp}(x)  \ = \  (1 - i)  \sqrt{x} \ + \ (1 + i)\ x^{3/2}  \ + \ O(x^{5 \over 2})  \ \ .
\end{equation} 
%
The expansion for large arguments (strong coupling expansion) is 
%
\begin{equation}
J_{\rm imp}(x) \ = \ 1  \ - \ {i  \over 4 x}  \ - \ {3  \over 32 x^2}  \ + \ 
O\left(x^{-3}\right)  \ \ .
\end{equation}
%
Interestingly, bulk edge coupling not only reduces the visibility of interference oscillations but also contributes a 
phase shift of $ \pi /4$ between the weak and strong coupling limit. This universal phase shift as a function of voltage is 
a signature of bulk-edge coupling in a non-abelian interferometer. 
The phase $e^{- i \pi/4}$ characteristic of the weak coupling limit
can be interpreted as the non-abelian part of the phase acquired by
two charge $-e/4$ QHs encircling each other in the clockwise
direction. This phase factor agrees with that obtained from the CFT
correlation function of two $\sigma$-operators, which describe the
non-abelian part of charge $\pm e/4$ particles. Alternatively, the
non-abelian phase can be inferred from the fact that non-abelian QHs
behave relative to each other as bosons if they are in the I fusion
channel, i.e.~that abelian and non-abelian part of the phase cancel
each other\cite{NaWi96,MoRe91,TsSi03}. Hence, the non-abelian phase has to
compensate the abelian phase $e^{i \pi/4}$ found from the operator
product expansion of two $e^{i \varphi(x)/2}$ QH operators.

\section{Lattice version of the continuum model}

In order to evaluate neutral correlation functions in the presence of bulk-edge coupling beyond perturbation theory, we 
develop a lattice description of the continuum model introduced in section II. For the lattice model, the parity expectation value 
 can be evaluated numerically for arbitrary strength of bulk edge coupling \cite{RoHaSiSt08}, 
and in section V we will derive the exact solution Eq.~(\ref{twoimpneutral.eq}) by using the inversion formula \cite{TrSc} for so-called
Hilbert type matrices. We shall first concentrate on the equal time correlation function 
%
\begin{equation} 
\langle {\cal N}_L {\cal N}_R \rangle \ = \ \langle P_{\rm edge}(-b/2,b/2) \rangle \ \equiv \  \langle P_b \rangle \ \ , 
\end{equation}
%
where $P(-b/2,b/2)$ is the edge parity operator for the up and down Majorana modes as
defined in Eq.~(\ref{edgeparity_static.eq}).
As a first step towards defining a lattice version of this operator, we consider a one-dimensional model of complex lattice fermions 
defined by the Hamiltonian
%
\begin{equation}
H_{\rm kin} \ = \ - {v_n \over 2 a} \sum_{j}  \Big( c_{j+1}^\dagger c_{j} + c_j^{\dagger} c_{j+1} \Big) \ \ ,
\end{equation}
%
where the sum runs over lattice points $j$, and the lattice constant is denoted by $a$. The kinetic energy describes a  dispersion relation 
%
\begin{equation}
\epsilon(k) \ = \ - {v_n \over a} \cos k  \ , \  \ \ -{\pi } < k < {\pi } \ \ .
\end{equation}
%
We study the model at half filling with $k_F = {\pi \over 2 }$. Defining $N = {b\over a}$,  the equal time parity expectation value
for the complex lattice fermions is given by
%
\begin{eqnarray}
  \langle  e^{i \pi \sum_{j=- N/2}^{N/2}  c_j^\dagger c_j}\rangle 
 & =  & \langle \prod_{- {N\over 2} < j < {N\over 2 }} (2 c^\dagger_j c_j -1) \rangle \ \ .
 \label{parity_complex.eq}
\end{eqnarray}
%
It is expressed as the expectation value of  a product of lattice operators. 
In order to make contact with a non-abelian 5/2-edge, we define Majorana operators 
%
\begin{subequations}
\begin{eqnarray}
\gamma_{j} & = & e^{i j \pi/2} \ c_j \ + \ e^{- i j \pi/2}\ c_j^\dagger \\
\tilde{\gamma}_{j} & = & {1 \over i} \big( e^{i j \pi/2} \ c_j \ - \ e^{- i j \pi/2}\ c_j^\dagger
\big)  \\
 i \gamma_{j} \tilde{\gamma}_{j} & = & 2 c_j^\dagger c_j -1 \ \ . \label{parityrelation.eq}
\end{eqnarray}
\end{subequations}
%
This transformation corresponds to a boost to the right moving Fermi point. The left moving Fermi point 
now corresponds to the momentum $ - \pi$. 
Using the equality Eq.~(\ref{parityrelation.eq}), the parity expectation value Eq.~(\ref{parity_complex.eq}) can be expressed 
as the expectation  value of a product of $\gamma$ and $\tilde{\gamma}$  
operators. As the Hamiltonian is quadratic, Wicks's theorem can be used to evaluate the expectation value of this product as the 
Pfaffian form of the matrix of correlation functions.  The Hamiltonians for the $\gamma$- and $\tilde{\gamma}$-mode are
%
\begin{subequations}
\begin{eqnarray}
{\cal H} & = & {v_n \over 4 a} \int_{- \pi}^{\pi} {d k \over 2 \pi} \gamma_{-k}
\ \sin(k ) \; \gamma_{k}\ \ , \\
\tilde{\cal H} & = & {v_n \over 4 a} \int_{- \pi}^{\pi} {d k \over 2 \pi} \tilde{\gamma}_{-k}
\  \sin(k ) \; \tilde{\gamma}_{k} \ \ .
\end{eqnarray}
\end{subequations}
%

As there is no coupling term between the two Majorana modes, the correlation function matrix decomposes into a block with  $\gamma$--correlators and another block with $\tilde{\gamma}$--correlators. Although we initially need both modes to write down an expression for the local parity operator,   the determinant of the correlation function matrix factorizes and it is sufficient to consider the $\gamma$-Majorana mode only. We can now identify the 
right-moving branch of $\gamma$ with the upper edge and the left-moving branch with the lower edge.

States with momenta between $-  \pi $ and $0$ are occupied, hence the zero temperature correlation function is given by
%
\begin{equation}
 \langle i \gamma_{j} \gamma_{l} \rangle \ = \ 2 i  \int_{- \pi}^0
{d k \over 2 \pi} \ e^{i k (j - l) }\ =
 {1   \over \pi}   \ {\big[1 - (-1)^{j - l} \big] \over j - l}  \ \ .
 \label{correlation_clean.eq}
\end{equation}
%
The correlation function vanishes  if $j - l$ is even, and is odd under exchange of $j$ and $l$. The parity expectation value for a system of 
a right moving and a left moving Majorana mode is given by
%
\begin{eqnarray}
\langle P_{\rm b} \rangle   & = &  \langle \prod_{- {N \over 2} < j < {N \over 2}}  \big(\sqrt{i } \gamma_j \big)\rangle \ = \  
 \sqrt{{\rm det}(\langle i  \gamma_j \gamma_{j^\prime} \rangle)}
\ \ . \label{squaredet.eq}
\end{eqnarray}
%
The r.h.s.~of Eq.~(\ref{squaredet.eq}) is positive  as required because the eigenvalues of the matrix of correlation functions  $\langle i  \gamma_j \gamma_{j^\prime} \rangle$ occur in pairs $\pm i  \lambda$ with real $\lambda$. 
When evaluating the determinant for different system sizes numerically, 
on finds that it  decays as  $N^{-1/4}$ in agreement with the analytical result obtained by using 
$\sigma$--correlators in the Ising CFT. 

We next  want to calculate the influence of localized bulk modes on the parity expectation value. More specifically, we need to calculate the expectation value
%
\begin{equation}
\langle P_{b, \rm imp} \rangle \ = \ \langle i \Gamma_u \Gamma_d \prod_{- {N\over 2} < j < {N \over 2}}  \big(\sqrt{i } \gamma_{j} \big)\rangle \ \ .
\label{imp_paritydefinition.eq}
\end{equation}
%
In order to evaluate this expectation value, we need to know the edge-edge, the impurity-impurity,  and the impurity-edge correlation functions in the presence of 
a coupling between impurities and edge. 
The lattice version of the bulk-edge coupling Eq.~(\ref{bulk-edge-tunneling}) is
%
%
\begin{equation}
{\cal H}_{\rm scatter} \ = \ \int {dk\over 2 \pi} \ f(k) \big[ i {\lambda_u \over \sqrt{a}} \Gamma_u \gamma_k \ + \ i {\lambda_d \over \sqrt{a}} \Gamma_d \gamma_{-2 k_F + k} \big] \ \ .
\end{equation}
%
Here, $f(k) = f(-k)$ is unity for momenta $k \ll 1 $ and drops rapidly to zero for larger momenta, such that 
 the dispersion relation can be linearized around the two Fermi points. As the Hamiltonian for bulk and edge Majorana 
 states is quadratic, all correlation functions needed for the evaluation of the parity expectation value Eq.~(\ref{imp_paritydefinition.eq}) can be evaluated exactly. After integrating out the edge Majorana modes, one obtains 
 an effective action for the bulk states
%
\begin{widetext}
\begin{eqnarray}
{\cal S}_{\rm imp} \! &   =&\!   {T \over 2}  \sum_{\epsilon_n} \left[
\Gamma_{u}(-\epsilon_n)\! \left( \! 
 - {i \epsilon_n \over 2} + \lambda_u^2 \! \int {dq \over 2 \pi} f(q) G_0(i \epsilon_n, q) \! \right)  \!\Gamma_u(\epsilon_n)
  +\Gamma_d(- \epsilon_n)\!
\left(\! - {i \epsilon_n \over 2} + \lambda_d^2 \! \int {dq \over 2 \pi} f(q) G_0(i \epsilon_n, q - 2 k_F)\! \right)\! \Gamma_d(\epsilon_n)\! \right] \ \ . \nonumber \\  \hspace{-1cm} & \hspace{-1cm} &   \hspace{-1cm}
\label{S_imp.eq}
\end{eqnarray}
\end{widetext}
%
Here, 
%
\begin{equation}
G_0(i \epsilon_n, k) \ = \ {2 \over i \epsilon_n - {v_n \over a} \sin(k )}
\end{equation}
%
denotes the edge Green function in the absence of scattering, $\epsilon_n = (2 n +1) k_B T/\hbar$ is a fermionic Matsubara frequency, and $T$ denotes temperature. Although we use a finite temperature formalism here, we 
will focus on the zero temperature case in the end. 
As there is no coupling between the bulk Majorana states,  the impurity-impurity correlator vanishes
$\langle i \Gamma_u \Gamma_d\rangle \equiv 0$. The correlation functions between impurity and edge operators are given by
%
\begin{eqnarray}
\langle i \Gamma_u(- \epsilon_l)  \gamma_k(\epsilon_l)\rangle & = & 
G_0(i \epsilon_l,k)\ f(k)\ {\lambda_u \over \sqrt{a}}\\
& & \times  \langle \Gamma_u(- \epsilon_l)\Gamma_u(\epsilon_l)\rangle_{{\cal S}_{\rm imp}}\nonumber \ \ , \\
\langle i \Gamma_d(- \epsilon_l)  \gamma_k(\epsilon_l)\rangle & = & 
G_0(i \epsilon_l,k- 2 k_F)\ f(k)\ {\lambda_d \over \sqrt{a}}\nonumber \\
& & \times  \langle \Gamma_d(- \epsilon_l)\Gamma_d(\epsilon_l)\rangle_{{\cal S}_{\rm imp}} \nonumber \ \ . 
\end{eqnarray}
%

Due to chirality, the edge-edge correlation function $\langle i  \gamma_l \gamma_{l^\prime}\rangle $ depends on the bulk-edge coupling strengths $\lambda_d$, $\lambda_u$ only if the coupling to the impurity occurs between the two lattice sites $l$ and $l^\prime$. As only  
 edge-edge correlation functions with both $l$ and $l^\prime$  inside the interferometer cell are needed  for the evaluation of 
 Eq.~(\ref{imp_paritydefinition.eq}),  these edge-edge correlators  become independent of the bulk-edge coupling strengths for $x_0 = b/2$, i.e.~for impurities coupling to the 
boundary of the interferometer cell. In order to simplify the task of calculating the full neutral correlation function 
including bulk-edge coupling, we will adopt $x_0 = b/2$ in the following. 
 

In order to extract the universal  long distance behavior of correlation functions, 
we linearize the dispersion relation around the two Fermi points and 
remove the momentum cutoff when possible. In this way, the Fourier transform $f_j$ of a function $f\left[{v_n \over a} \sin k\right]$ becomes
%
\begin{eqnarray}
f_j & = & \int_{- 3 \pi/2}^{\pi/2} {d k \over 2 \pi} e^{i k j}  f\left[{v_n \over d} \sin k\right] \nonumber  \\
& \approx &\left( - e^{i \pi j} + 1\right) \int_{-\infty}^\infty {d k \over 2 \pi} f\left[{v_n \over d} k \right] e^{i k j} 
\end{eqnarray}
%
The first  term with a rapidly oscillating position dependence is due to integrating over momenta $-3 \pi/2 < k - \pi/2$, whereas the second term with a smooth position dependence is due to integration over momenta 
$- \pi/2 < k < \pi/2$. With the help of this formula and regularizing momentum  integrals as ${i \over \pi} \int_0^\infty d k\, e^{i k (x + i \eta) } = {1 \over \pi }{1 \over x + i \eta}$, 
expression Eq.~(\ref{correlation_clean.eq}) is easily reproduced in the  limit $\eta \to 0$.
For the bulk-edge correlation functions, one finds 
%
\begin{eqnarray}
 \langle i \Gamma_{u} \gamma_j \rangle\! & = & \!  
{ 2 \lambda_{u} \sqrt{a} \over \pi v_n} \; e^{2 \lambda_{u}^2 j a /v_n^2} E_1(2 \lambda_{u}^2 j a /v_n^2) \ \ ,\\
\langle i \Gamma_{d} \gamma_j \rangle \! & = &\!  (-1)^{j}\;  
{ 2 \lambda_{d} \sqrt{a} \over \pi v_n } \; e^{2 \lambda_{d}^2 j a /v_n^2}  E_1(2 \lambda_{d}^2 j a /v_n^2)  .
\label{edgeimp_ana.eq}
\end{eqnarray}
%

Here, the exponential integral is defined as
%
\begin{eqnarray}
E_1(x) & = & \int_1^\infty dt \; {e^{- t x} \over t}  \ \ .
\label{expint.eq}
\end{eqnarray}
%
It has the asymptotic expansions
%
\begin{eqnarray}
E_1(x) & \to & - \gamma - \ln x \ \ \ \ {\rm for } \ \ \ x \to 0 \ \ ,\\
E_1(x) & \to & {e^{-x} \over x} \ \ \ \ {\rm for } \ \ \ x \to \infty \nonumber 
\end{eqnarray}
%
with  $\gamma = 0.57721...$ denoting Euler's constant. Although it is not needed for the calculations presented in this manuscript, we would like to mention the result for the 
edge-edge correlation function in the presence of bulk-edge coupling, which was used to 
obtain the numerical result for the reduction factor in Ref.~\onlinecite{RoHaSiSt08}. It is 
given by
%
\begin{eqnarray}
\langle \gamma_j \; \gamma_{l}\rangle_{\rm imp} & = &
{1 \over \pi} \ {1 \over j - l}\ [ 1 - (-1)^{j - l}] \\
& & \hspace*{-0.9cm}
- {4 \lambda_u^2 a\over \pi v_n^2} \ e^{2 \lambda_u^2 (j - l)a/v_n^2}\ E_1\big[2 \lambda_u^2 (j-l)a/v_n^2\big]\ \nonumber \\
& &\hspace*{-.9cm} \times \left[\Theta(j) \Theta(-l)
- \Theta(-j) \Theta(l)\right]
 \nonumber \\
& & \hspace*{-0.9cm}
-  (-1)^{j - l} \ {4 \lambda_d^2 a\over \pi v_n^2} \ e^{2 \lambda_d^2 (l -j)a/v_n^2}\ 
E_1\big[2 \lambda_d^2 (l - j)a/v_n^2\big]\ \nonumber \\
& & \hspace*{-.9cm}
\times \left[ \Theta(-j) \Theta(l) - \Theta(j) \Theta(- l)\right] \ \ .\nonumber
\label{edgecorrelation_ana.eq}
\end{eqnarray}
%

\section{Exact solution for parity correlation function with impurities}

As explained in the last section, we consider a geometry where the bulk impurity couples to the edge at the boundary of the 
interferometer cell, i.e.~$x_0= b/2$. This geometry somewhat simplifies calculations  because now the edge-edge correlation function does not have an impurity contribution. For ease of notation,  we assume ${a \lambda_u^2 \over 
v_n^2} = {a \lambda_d^2 \over  v_n^2} \equiv \lambda^2$ in the following, the generalization to two different couplings is 
straightforward.  

Again we will be calculating a correlation function by using Wick's
theorem to rewrite that correlation function as a determinant analogous
to Eq.~(\ref{squaredet.eq}).  However, here we will calculate the more complicated
correlation function Eq.~(\ref{imp_paritydefinition.eq}). 
We denote the matrix of correlation functions, whose determinant needs to be calculated, by ${\bf \sf C}$.  All diagonal elements of ${\bf \sf C}$ vanish. We adopt a bra-ket notation in the following and denote a position along the edge  by $|j\rangle$, and 
the two bulk impurities by $|u \rangle$, $|d \rangle$. The edge-edge correlation function is then given by
$\langle j|{\bf \sf C} |j^\prime \rangle$, the bulk-edge correlation is 
%
\begin{eqnarray}
\langle j|{\bf \sf C} |u \rangle & = &  { 2 \lambda \over \pi} e^{2 \lambda^2 j} E_1(2 \lambda^2 j) \ \ ,  \nonumber \\
\langle j|{\bf \sf C} |d  \rangle & = & (-1)^j {2 \lambda \over \pi} e^{2 \lambda^2 j} E_1 (2 \lambda^2 j)  \ \ \ .
\end{eqnarray}
%
The impurity-impurity correlation function is $\langle u | {\bf \sf C}|d\rangle = \langle d | {\bf \sf C}|u\rangle
\equiv 0$. Our result
is again a square root of a determinant, and the determinant
 is the product of all eigenvalues. Let us 
assume we know the eigenvalues of the edge-edge part. 
For small $\lambda$, the bulk-edge correlation functions $\langle j|{\bf \sf C} |u \rangle $, $\langle j|{\bf \sf C} |d  \rangle$  are of order 
$\lambda \ln^2 \lambda$, and the leading contribution to the determinant is obtained by multiplying the determinant of 
the edge-edge correlators with the perturbatively calculated eigenvalues of the impurity-impurity part of the matrix. 
Although the eigenvalues are calculated perturbatively in $\lambda$, which is proportional to the lattice constant 
and goes to zero in the continuum limit, the final result is valid even in the strong coupling regime with 
large $\lambda N$.  The square root of the product of these two eigenvalues is the reduction factor $R(b)$, i.e.~the ratio 
%
\begin{equation}
R \ = \  {\langle {\cal N}_L {\cal N}_R i \Gamma_u  \Gamma_d  \rangle \over 
 \langle {\cal N}_L {\cal N}_R  \rangle_0}
\end{equation}
%
of the neutral expectation value in the presence of two impurities to the expectation value without
impurities. 

 Without bulk-edge coupling, {\bf \sf C} has two 
zero eigenvalues.  To determine the shift of these zero eigenvalues due to the coupling between impurities and 
edges, we use second order perturbation theory to calculate the effective matrix elements $\langle u|{\bf \sf C}|d \rangle_{\rm eff}$ due to "virtual transitions" of a bulk Majorana to the edge and back. Up to a sign,  the reduction factor is then 
equal to this effective matrix element, 
%
\begin{equation}
R \ = \ \langle u|{\bf \sf C}|d \rangle_{\rm eff} \ \ .
\end{equation}
%
To  calculate the effective matrix element, we change to a new 
basis
%
\begin{equation}
|a\rangle = |u\rangle + |d\rangle \ \ , \ \ \ \  |b\rangle = |u \rangle - |d \rangle \ \ .
\end{equation}
%
In the new basis, $|a\rangle$ only couples to even lattice sites, while $|b\rangle$ only couples to odd ones. To 
exploit this, 
it is useful 
to decompose the lattice into even and odd sites according to 
%
\begin{equation}
j = 2 n + {1 \over 2} ( 1 + \sigma) \ \ , \ \ \ \ \  j^\prime = 2 n^\prime + {1 \over 2}(1 + \sigma^\prime) \ \ .
\label{nsigma.eq}
\end{equation}
%
We assume that the number of lattice sites $N$ is even such that this decomposition works. Then, $n$ runs from $0$ to $N/2$, and $\sigma = \pm 1$ determines whether the lattice site is even or odd. In analogy to Eq.~(\ref{nsigma.eq}), we use an $|n \sigma\rangle$ basis in the following, where $|n \sigma\rangle \equiv |j\rangle $ with $j$ given by Eq.~(\ref{nsigma.eq}).
Denoting the eigenvectors of the bulk-bulk part {\bf \sf C}$^0$  of {\bf \sf C} by $|{\bf \sf e}_l\rangle$ and the corresponding 
eigenvalues by $\lambda_l$, the effective matrix element between $| a \rangle$ and $| b \rangle$   state is given by
%
\begin{eqnarray}
\langle b|{\bf \sf C}|a\rangle_{\rm eff} & = & \sum_{l=1}^N {\langle b | {\bf \sf C} | {\bf \sf e}_l\rangle 
\langle {\bf \sf e}_l| {\bf \sf C} | a \rangle \over \lambda_l}\label{matrixeffective.eq} \\
& = & \sum_{j,j^\prime} \langle b | {\bf \sf C} | j\rangle \langle j | ({\bf \sf C}^0)^{-1} | j^\prime \rangle \langle j^\prime
| {\bf \sf C}| a \rangle \nonumber \\
& = & \sum_{n n^\prime} \langle b | {\bf \sf C} | n +\rangle \langle n + | ({\bf \sf C}^0)^{-1} | n^\prime - \rangle \langle n^\prime
- | {\bf \sf C}| a \rangle \nonumber 
\end{eqnarray}
%
with (see Eq.~(\ref{edgeimp_ana.eq})
%
\begin{equation}
\langle b | {\bf \sf C} | n +\rangle = \langle n
- | {\bf \sf C}| a \rangle = { 4 \lambda \over \pi} e^{4 \lambda^2 n} E_1\left(4 \lambda^2 n\right) \ \ .
\end{equation}
%
The physical interpretation of the effective matrix element Eq.~(\ref{matrixeffective.eq}) is that the $|a\rangle$ bulk state makes a virtual transition to the edge and then back to the 
$|b\rangle$ state. For this reason, the double sum in  the second  line of Eq.~(\ref{matrixeffective.eq}) runs over edge states  only.  
In order to calculate matrix elements of $({\bf \sf C}^0)^{-1}$, we use the fact that in the $n$, $\sigma$ basis $ {\bf \sf C}^0$
has the form
 %
 \begin{eqnarray}
 {\bf \sf  C}^0 = \left( 
 \begin{array}{cc}  {\bf \sf 0} & {\bf \sf D } \\ - {\bf \sf D}^T & {\bf \sf 0 }  \end{array}
 \right)
 \end{eqnarray}
 %
with 
%
\begin{equation}
D_{n n^\prime} = C^0_{n n^\prime, - +} = {2 \over \pi} {1 \over 2 (n - n^\prime) - 1} \ \ , 
\end{equation}
%
see Eq.~(\ref{correlation_clean.eq}). Then, 
%
\begin{equation}
 \langle n + | {\bf \sf C}^{-1} | n^\prime - \rangle  = \left( {\bf \sf D}^{-1}\right)_{n n^\prime} \ \ .
\end{equation}
%
Matrices of the
form of {\sf D} are known as Hilbert-type, and using the inversion formula derived by Trench  Scheinok \cite{TrSc} we find that the  inverse of  {\rm \sf D} is given by
%
\begin{eqnarray}
 \left( {\bf \sf D}^{-1}\right)_{m n}&  = & {\pi \over 4} {1 \over n - m - {1 \over 2}} \prod_{q \neq m} \left[ 1 - {1 \over 
 2(q - m)}\right] \\
 & & \times \prod_{s \neq n} \left[ 1 - {1 \over 2 (n - s)}\right] \nonumber \\
 & = & {1 \over \pi} {1 \over n - m - {1 \over 2}} {\Gamma\left({1 \over 2} + m\right) \Gamma \left({1 \over 2} + {N \over 2} - m\right)
 \over \Gamma(m) \Gamma\left(1 + {N\over 2} - m\right)} \nonumber \\
 & & \times  {\Gamma\left(n - {1 \over 2}\right) \Gamma\left({3 \over 2} + {N\over 2} - n \right)
 \over \Gamma(n) \Gamma\left(1 + {N\over 2} - n \right)} \ \ . 
 \end{eqnarray}
 %
As we will finally take the limit $N\to \infty$ with $N \lambda^2$ fixed, we can use Sterling's formula to simplify
%
\begin{eqnarray}
 \left( {\bf \sf D}^{-1}\right)_{m n} \approx { 1 \over \pi}  {1 \over n - m - {1 \over 2} } \sqrt{m \over n} \sqrt{1 + {N\over 2} - n \over 1 + {N \over 2} - m} \ \ .
\end{eqnarray}
%
As the formula Eq.~(\ref{matrixeffective.eq}) uses only the symmetric part of {\bf \sf D}$^{-1}$, we symmetrize and obtain
%
\begin{equation}
\left( {\bf \sf D}^{-1}\right)_{{\rm even},  m n} = - {N +2 \over 4 \pi}  {1 \over \sqrt{m n} \sqrt{(1 + {N\over 2}  -n) ({N\over 2} +1 -m}} \ \ .
\end{equation}
%
Taking everything together, the reduction factor is 
%
\begin{equation}
R = {N + 2 \over 4 \pi} \left({ 4 \lambda \over \pi}\right)^2 \left[ \sum_{n-1}^{N \over 2} e^{4 \lambda^2 n} E_1\left(4 \lambda^2 n\right) {1 \over \sqrt{n} \sqrt{{N\over 2} +1 - n}} \right]^2 \ . 
\end{equation}
%
Taking the continuum limit, one finds
%
\begin{eqnarray}
R & =  & \lambda^2 N {2 \over \pi^3} \left[ \int_0^{\lambda^2 N} dx e^{2 x} E_1\left(2 x\right) {1 \over \sqrt{x (\lambda^2 N - x)}}
\right]^2 \nonumber \\
& = & \lambda^2 N {2 \over \pi} \left[ e^{\lambda^2 N} K_0\left(\lambda^2 N\right) \right]^2 \ \ . 
\label{reduction_exact.eq}
\end{eqnarray}
%
For the evaluation of the integral in the last equation, we used the integral representation 
Eq.~(\ref{expint.eq}) for $E_1(x)$  and evaluated the $x$-integral in terms of a modified Bessel function 
$I_0(x)$, for details see Ref.~\onlinecite{GraRy}. The remaining integral is again tabulated in 
Ref.~\onlinecite{GraRy}. One sees that the reduction factor is the square of reduction factors due to the two individual impurities. Using the asymptotic behavior of the zeroth order modified Bessel function $K_0(z) \approx - \ln z$ for $z \to 0$ and 
$K_0(z) \approx \sqrt{\pi \over 2 z} e^{-z}$ for $ z \to \infty$, we find the asymptotic behavior of the reduction factor
%
\begin{eqnarray}
R(\lambda^2 N \ll 1)& = & {2 \over \pi} \lambda^2 N \left( \ln \lambda^2 N\right)^2 \nonumber \ \ , \\
R(\lambda^2 N  \to \infty) & = & 1 \ \ . 
\end{eqnarray}
%
Because of the chirality of upper and lower edge,  the time dependence of the exact solution can be obtained by 
replacing $b \to b - i \tau$ in the factor describing  the upper edge  and $b \to b + i \tau$ in the factor describing  the lower edge in
Eq.~(\ref{reduction_exact.eq}) such that the time dependent reduction factor is given by
%
\begin{eqnarray} 
R(b, \tau) & = &   {\lambda_u \lambda_d \over v_n^2} \ {2 \over \pi} \ \sqrt{b^2 + v_n^2 \tau^2}
\label{reduction_time.eq}\\
& & \times \left[ e^{\lambda_d^2 (b + i v_n\tau)/v_n^2} K_0\left(\lambda_d^2 (b+ i v_n \tau)/v_n^2 \right) \right]\nonumber \\
& & \times 
\left[ e^{\lambda^2_u (b - i v_n \tau)/v_n^2} K_0\left(\lambda_u^2 (b - i v_n \tau_/v_n^2\right) \right] \nonumber  \ \ . 
\end{eqnarray}
%
 This extension of the static solution to 
finite time differences can be justified by considering the case of one bulk impurity coupled to one of the 
edges, say the upper one, first. Then, we  define a neutral correlation function
%
\begin{equation}
{\cal N}_u(x_2,\tau_2;x_1, \tau_1; x_0) \  = \ \langle T_\tau \sigma_u(x_2, \tau_2) \sigma_u(x_1,\tau_1) \sqrt{i} \Gamma_u \rangle \ ,
\end{equation}
%
which in principle could depend on the four arguments $x_1$, $x_2$, $\tau_1$, $\tau_2$ and the parameter $x_0$ (position of bulk-edge coupling) separately. As the impurity is static, there is translational invariance in time and ${\cal N}_u$ can only depend on the difference $\tau_2 - \tau_1$. 
As we consider  a situation where the impurity is coupled to a point inside the cell, we have $x_2 < x_0$. 
Then, due to chirality, the field  $\sigma_u(x_2,\tau_2)$ is not influenced by bulk-edge coupling and  does not depend on $x_2$ and $\tau_2$ separately but only on the combination $x_2 + i \tau_2$, and it satisfies the differential equation
$(\partial_{x_2} + i \partial_{\tau_2}) \sigma_u(x_2, \tau_2) = 0$. If we restrict ourselves to time differences
$\tau_2 - \tau_1 \neq 0$ different from zero, the correlation function ${\cal N}_u$ satisfies the same 
differential equation and can for this reason only depend on the variables $x_2 + i(\tau_2 - \tau_1)$ and 
$x_1$.  However, from Eq.~(\ref{reduction_exact.eq}) we see that the static correlation function 
only depends on the difference $x_1 - x_2$, so we can conclude that 
${\cal N}_u(x_2,\tau_2;x_1, \tau_1; x_0)$ is a function of the single variable 
$x_1 - x_2 - i (\tau_2 - \tau_1)$, and that the correct analytic continuation of Eq.~(\ref{reduction_exact.eq})
is indeed given by replacing $b \to b - i \tau$ with $b = x_1 - x_2$ and $\tau = \tau_2 - \tau_1$. 
The analytic continuation for the lower edge can be derived by a similar argument.

\section{Interpretation in terms of resummed perturbation theory}

In this section, we show that the exact solution Eq.~(\ref{reduction_exact.eq}) can be reproduced by 
resumming the  perturbative expansion of the neutral correlation function in powers of 
the bulk-edge coupling constant. The terms 
contributing to  this resummation are those which turn the zeroth 
order bulk-bulk correlation function $\langle T_\tau \Gamma_r (\tau) \Gamma_r(0)\rangle_0 \equiv 1$
into the full correlator.  The $\sigma_u \sigma_u \psi_u$-correlation function appearing in the lowest order expression Eq.~(\ref{neutral_perturbative.eq}) is not modified 
in the perturbative expansion due to the special choice  $x_0 = b/2$, which implies that  the chiral $\sigma_u \sigma_u \psi_u$-correlation function is only evaluated for spatial arguments 
 $x \leq x_0$. As one can see for example from the edge-Majorana  correlation function Eq.~(\ref{edgecorrelation_ana.eq}), bulk-edge coupling is only important if one spatial argument is to the left 
and another one to the right of $x_0$.  

Since upper and lower edge decouple in perturbation theory (modulo fusion channels),  we consider only one impurity coupled to one edge, say the upper edge. 
We start by calculating the neutral equal time correlation function 
%
\begin{equation}
{\cal I}_u \ = \ \langle \sigma_u(- b/2) \sigma_u(b/2) \sqrt{i} \Gamma_u \rangle 
\label{neutralup.eq}
\end{equation}
%
in perturbation theory. The lowest order contribution is \cite{RoHaSiSt08}
%
\begin{eqnarray}
{\cal I}_u^{(1)}  \!& = & \!  - \sqrt{i} \lambda_u \! \int_{- \infty}^\infty \! \! \! \! \! d\tau \langle T_\tau \sigma_u(- {b\over 2}, 0+) \sigma_u({b\over 2},0-)
 \psi_u({b\over 2},\tau)\rangle_0 \nonumber \hspace*{-1cm} \\
 & & \times \langle T_\tau \Gamma_u(\tau) \Gamma_u(0)\rangle_0 \ \ .
 \label{neutral_perturbative.eq}
\end{eqnarray}
%
This  expression is logarithmically divergent and needs to be regularized by a cutoff on the time integral, which  in Ref.~\onlinecite{RoHaSiSt08} was inserted by hand.
However, when resumming the infinite set of diagrams which turns the zero order correlator into the full correlator, 
the integral  becomes finite and we are able to reproduce the exact solution for the reduction factor in Eq.~(\ref{reduction_exact.eq}). In order to verify this proposition, we calculate the expression Eq.~(\ref{neutral_perturbative.eq}) with $\langle T_\tau \Gamma_u(\tau) \Gamma_u(0)\rangle_0$ replaced by the full correlation function $\langle T_\tau\Gamma_u(\tau) \Gamma_u(0)\rangle$. From the action Eq.~(\ref{S_imp.eq}) we find in frequency 
space
%
\begin{equation}
- \langle \Gamma_u(- \epsilon_l) \Gamma_u(\epsilon_l)\rangle \ = \ {2 \over i \epsilon_l + {2 i \lambda_u^2 \over v_n} {\rm sign} \epsilon_l} \ \ .
\end{equation}
%
After calculating the Fourier transform to Matsubara time we find
%
\begin{equation}
- \langle T_\tau \Gamma_u(\tau) \Gamma_u(0)\rangle  =  {1 \over \pi} {\rm Im} \left[ e^{- i  \tau  2 \lambda_u^2/v_n} 
E_1(- i \tau 2 \lambda_u^2/v_n)\right] \ \ .
\end{equation}
%
The calculation can be expressed in a more compact fashion by defining
%
\begin{equation}
g(-z) \ = \ {1 \over \pi}  e^{- 2 z \lambda_u^2/v_n} E_1(- 2 z \lambda_u^2/v_n) \ \ . 
\end{equation}
%
In addition, we make use of the CFT correlation function  
%
\begin{eqnarray}
\langle \sigma_u(-b/2,0+)  \sigma_u(b/2,0-) \psi_u(b/2,\tau)\rangle_0 & =&  {1 \over \sqrt{2 \pi }} \left(-b\right)^{3 \over 8} \nonumber \\
& & \hspace{-2cm}\times {1 \over \sqrt{(b + i \tau) i\tau}} \ \ .
\label{sigmasigmapsi.eq}
\end{eqnarray}
%
Note that the additional factor $1/\sqrt{\pi}$ in Eq.~(\ref{sigmasigmapsi.eq}) as compared to 
Ref.~\onlinecite{RoHaSiSt08} is due to 
the difference in the Majorana Lagrangian Eq.~(\ref{majorana.eq}) as compared to \onlinecite{RoHaSiSt08}.
Now we can express the equal time correlation function  as 
%
\begin{eqnarray}
{\cal I}_u  & = &  \sqrt{i} \lambda_u  (- b)^{3 \over 8}    {1 \over \sqrt{2 \pi }}  \int_{- i \infty}^{i \infty} {dz \over i} \  {g(-z) - g(z) \over \sqrt{  z(z + b)}}   \ . 
\end{eqnarray}
%
We  close the integration contour in the right half plane for the integral over $g(z)$ and in the left half plane for 
$g(-z)$. We are allowed to do so because $g(z) \sim 1/z$ for large $z$ in the right half plane, which together with the 
asymptotic $1/z$ behavior of the $\langle \sigma \sigma \Psi\rangle$ correlator makes sure that the infinite semicircle does not 
contribute to the integral. As the $\langle \sigma \sigma \Psi\rangle$ correlator has a cut only along the negative real axis  between  $z=-b$ and $z=0$, the integral over $g(z)$ vanishes. The integral over $g(-z)$ can be converted into  a contour 
encircling this cut and gives
%
\begin{eqnarray}
{\cal I}_u   & = &    
 \lambda_u \sqrt{i}  (- b)^{3 \over 8}  \sqrt{2 \over \pi^3} \int_0^b dx \ {e^{-2 x \lambda_u^2 /v_n} 
E_1(2 x \lambda_u^2/v_n) \over \sqrt{(b - x) x}}  \ \ . \nonumber \\
\label{Iu.eq}
\end{eqnarray}
%
We note that the expression for ${\cal I}_d$ differs from Eq.~(\ref{Iu.eq}) by a factor of 
$i$ due to the opposite chirality in the correlator Eq.~(\ref{sigmasigmapsi.eq}). 
For this reason, the reduction factor obtained from the product ${\cal I}_u {\cal I}_d$ 
is real in agreement with Eq.~(\ref{reduction_exact.eq}).
The expression  Eq.~(\ref{Iu.eq})  agrees up to a phase factor with the correlation function obtained from the square root of the 
reduction factor Eq.~(\ref{reduction_exact.eq}). As an additional benefit, 
 using the finite temperature expressions for the $\langle \sigma \sigma \Psi \rangle$ 
and $\langle \Gamma_u \Gamma_u\rangle$ correlators opens a route towards a generalization to finite temperature.

\section{Discussion}

In this paper, we have analyzed  the influence of a tunnel coupling between bulk and edge Majorana states  on the visibility and phase of 
interference oscillations in a non-abelian $\nu=5/2$ quantum Hall interferometer. Such a tunnel coupling is important 
because it blurs the distinction between bulk and edge degrees of freedom and thus complicates the  observation of  the even-odd effect as a signature for non-abelian statistics. In our discussion, we have focused on the behavior at temperature $T=0$, for an interferometer encircling  one or two localized quasiparticles (QPs),   as a function of the source-drain voltage and the strength of coupling between the Majorana modes of the impurities and the neutral modes of the edge.

The present paper is an extension of results presented in a previous letter by the authors.\cite{RoHaSiSt08}  In the present paper, we have found an exact analytic formula for the equal-time parity correlation function for the two ends of the interferometer, when the localized quasiparticles are both located close to one end, and we have verified that the correlation function saturates, in the strong coupling limit, at the same value as one would find in the absence of localized quasiparticles.   This correlation function was only obtained numerically in our previous work. 
Analyzing the analytic properties of the function in the space-time plane,
we now obtain an analytic form for the correlation function at two different times, and from that we can predict the dependence of the interference amplitude on the applied  voltage and the bulk-edge coupling strengths.   
We have also been able to examine the phase shift in the interference pattern introduced by the presence of a finite coupling between the bulk QPs and the edge.  In addition, the current  paper presents various details of the analysis that had to be omitted from Ref. \onlinecite{RoHaSiSt08} due to lack of space.

In our work, we have particularly examined the case of a short interferometer, or relatively low voltage, where the interference visibility is largest.   Specifically, we assume
$t_V \gg t_b$,  where, $t_b = b/v_n$ is the time needed for a neutral excitation  to move along one edge, from one constriction the other, and $t_V= \hbar /e^\star V$ is
the extension in time of a QP wave packet transferred from one edge to the other by backscattering at one of the constrictions.   
We shall summarize here the qualitative results of our studies, and then say a few words about their  implications for experiments.

We start by discussing the case of  a single 
bulk Majorana mode inside the interferometer. In the absence of bulk-edge coupling, the leading harmonic of the 
interference signal vanishes.  In agreement with  previous 
analyses of this problem \cite{overbosch07,RoHaSiSt08}, we find that for weak coupling, at $T=0$, interference can be observed but is reduced by a factor proportional to $\sqrt{t_V/t_\lambda}$, where $t_\lambda= (\pi \lambda^2/2 v_n)^{-1}$ is the  characteristic tunneling time associated with the exchange of a Majorana particle between the localized QP and the edge. 
For large values of the effective coupling constant $t_V/t_\lambda$, the bulk Majorana mode 
is effectively absorbed by the edge and the interference signal is fully restored, such that  the strong coupling case corresponds
to an interferometer with no bulk degree of freedom. In addition, on the way from weak to strong coupling, the phase of the interference signal is shifted by $ \pi/4$. Although bulk-edge coupling enforces a modification of the way one 
looks at the even-odd effect, it actually enriches this effect with a new direction in parameter space, as the dimensionless coupling strength $t_V/t_\lambda$ depends on source drain voltage. The signature for an odd number of impurities, with just one of them coupled to the edge, is not the  complete absence of the leading 
harmonic, but rather a reduced amplitude, which depends on the applied voltage. The interference intensity grows with decreasing voltage at a rate which is  enhanced relative to the behavior in the absence of a bulk Majorana mode (the intensity grows $\propto 1/V$ instead of $\propto 1/\sqrt{V}$ ) until the reduction factor saturates at unity.  In addition,  when the reduction factor is small compared to unity, the interference factor will have  universal phase shift of $ \pi/4$ relative to the pattern in the strongly coupled regime.   

Our results for a single bulk QP at  $T=0$ can be readily generalized to the case of finite temperatures.  Qualitatively, the temperature will make  relatively little difference as long as $kT$ is small compared to $e^* V$. However in the opposite limit, temperature will be important, and its effect may be roughly estimated by replacing the the voltage time $t_V$, in the formulas above, by the thermal time 
$t_T \equiv \hbar / kT$.  A more quantitative analysis can be given, but it will not be discussed here.  

In the case of two bulk Majorana modes coupled to opposite  edges,   with comparable coupling strengths, the average value of the interference signal shows the same qualitative behavior as in the case of a single impurity. 
However, the average value of the interference current now has a more subtle interpretation than in the case of a single 
impurity: due to the presence of the factor $i \Gamma_u \Gamma_d$ in the tunneling operator, the interference signal is sensitive to the state of the two bulk Majorana modes  even in the absence of 
bulk-edge coupling. Although in the absence of bulk-edge coupling the expectation value of $i \Gamma_u \Gamma_d$  does not change in time once it is prepared in an eigenstate, the quantum statistical average $\langle i \Gamma_u \Gamma_d\rangle $ is zero, as the average is taken over both possible states of the system. Experimentally, then, the quantum statistical average corresponds to a situation where the interference signal is averaged over different initializations of 
the bulk states.

In the presence of weak but non-zero coupling between the bulk and edge modes, two things happen:
i) the quantum mechanical average 
$\langle i \Gamma_u \Gamma_d\rangle  \propto  (t_v / t_\lambda) \ln^2[t_\lambda/ t_V]$ is now finite; and ii) for a single experiment starting with the 
bulk states initialized in an  eigenstate of $i \Gamma_u \Gamma_d$, the interference phase will fluctuate on the time scale 
$\min (t_{\lambda u} , t_{\lambda d})$,
so that the time average of the interference current over these 
quantum mechanical fluctuations is equal to the quantum statistical average.

Now we must distinguish between two experimental situations.  If the experimental measurement is averaged over a time $ t_{\rm{av}} $ that is long compared to one or both of  the switching times $ t_\lambda$, then the experiment will measure the statistical average interference signal, which will be very close to zero if $t_\lambda $ is much longer than $t_V$ and $t_T$.   On the other hand, if the bulk modes are so weakly coupled to the edge that  for both bulk QPs, $ t_\lambda > t_{\rm{av}}$, then the experiment will not measure a statistical average,  but will see only one or the other of two possible fermion states.   In this case one will see a full interference signal, with the same amplitude as if the impurities were not there at all, but with a phase that depends on the starting configuration of the system.   

More generally, we may distinguish three ranges of coupling strengths for a QP localized in the bulk.  If coupling to the edge is so weak that $ t_\lambda > t_{\rm{av}}$, we may say that the bulk state is ``decoupled" from the edge.  If the coupling is in the range where $ t_\lambda$ is small compared to $t_V$ and $t_T$, then QP is ``strongly coupled" to the edge.  If the coupling strength is in the range where $ t_\lambda$ is small compared to $t_{\rm{av}}$, but large compared to either $t_V$ or $t_T$, then we may say that the QP is ``weakly coupled" to edge. 

In an experiment where the interfering edge encloses two or more localized QPs, we may ignore any strongly coupled QPs, as they will be  effectively incorporated  into the edge. If there are no ``weakly coupled" QPs inside the loop, then the presence or absence of the interference signal corresponding to charge e/4 will be determined by the number of decoupled QPs inside the loop.  The interference pattern will be present if this number is even, and absent otherwise.  If there are any weakly coupled  QPs inside the loop, however, with relaxation time $t_\lambda$  small compared to the experimental averaging time, then the interference pattern will be absent, regardless of the number of decoupled QPs that may also be present.

In a recent experiment by Willett et al.~\cite{Willett08}, resistance oscillations in a Fabry-Perot device were studied experimentally. For magnetic fields near a bulk filling fraction
$\nu=5/2$, oscillations in the longitudinal resistance were observed as  a function of  side gate voltage.
Depending on the range over which the side gate voltage was varied,
consecutive doubling and halving of the voltage period of resistance oscillations was observed. The side gate voltage  was interpreted as changing the area of the interferometer cell.
If the bulk filling fraction deviates slightly from the exact value of $5/2$, a change of area will once in a while change  the number of QPs inside the interferometer cell by one. If the number
of QPs changes from even to odd, the fundamental harmonic is suppressed and the
period of the interference signal  is halved, while for a change from an odd to an even number of localized QPs inside the cell, the period is doubled. A reduced voltage period in the presence 
of an odd number of localized QPs can  arise from interference of abelian charge $\pm e/2$ QPs or from  non-abelian charge $\pm e/4$ QPs encircling  the interferometer cell twice. This interpretation of
the experiment~\cite{Willett08} is discussed in more detail in Ref.~\onlinecite{Qguys09}. 

A difficulty with this interpretation is that it seems to require
that all  QPs  inside the edge are either completely decoupled
from the edge, or strongly coupled so that they are incorporated
into the edge.  If there any QPs in the intermediate weakly
coupled range, where the Majorana state changes back and forth
many times during the averaging time of the experiment,  then  the
interference corresponding to charge e/4 would be completely
absent, or at least greatly reduced in size.  It is not clear to
us, why there should be no weakly coupled QPs inside the
interferometer in the interference experiments of Ref.
\onlinecite{Willett08}, nor is it clear why one should be rid of
their effects if such QPs are present.  We do note, however, that
one possible ingredient that is missing from our analysis is a
tunnel coupling that allows for an exchange of Majorana fermions
between bulk quasi-particles. In the presence of such a coupling,
the degeneracy of the localized bulk Majorana states is removed.
The resulting spectrum is then composed of several states. Each of
these states corresponds to an interference pattern, whose
amplitude and phase are determined by the expectation value and
fluctuations of the parity operator in that state. The coupling of
the bulk states to the propagating Majorana mode of the edge
introduces a width to these states, equilibrates them to the
temperature of the edge, and thermally averages the interference
pattern. Then, if the splitting between the bulk states is large
compared to their width and to the temperature, a well defined
interference pattern will be observed. A recent estimate \cite{Baraban+09} for the tunnel splitting between quasi-particles is rather
sizeable, about $100mK$ for a separation of $0.1$ micron. This
effect and the overall problem clearly require further
investigation.

Altogether, the experimental results obtained so far are not yet well understood by theory, and it is not yet clear that the observed effects originate from the unique properties
 of non-abelian
quasi-particles at $\nu=5/2$. It is possible that a clue to that
question may be obtained from the transition region between two
different periods, when a change in the number of localized QPs
happens due to a change in area of the interferometer cell, and when the
distance between edge and bulk QP decreases as a function of
side gate voltage. Then, the bulk-edge coupling should change from
small to large values, and one can expect that  the theory
developed in this manuscript is applicable. It would be
interesting to analyze the data of Ref.~\onlinecite{Willett08} from the
point of view of bulk-edge coupling, and to test the theoretical
prediction that in the transition region between different gate
voltage periods  the interference current has a modified power law
dependence on source-drain voltage and that there is a  $\pi/4$
phase shift as a function of source-drain voltage.

Acknowledgments:  After completion of this work we learned that a similar  problem was solved 
by W.~Bishara and C.~Nayak using a different method \cite{BiNa09}. We 
would like to thank  W.~Bishara and C.~Nayak  for interesting discussions and for sharing unpublished results with us. We are indebted  to  V. Gritsev for leading us to Ref.~\onlinecite{TrSc}. This work was supported in part by 
the Heisenberg program of DFG, by NSF grant DMR 0541988, by a grant from the Microsoft Corporation,  by the US-Israel Binational Science Foundation, and the Minerva foundation.

\end{document}